\documentclass[onecolumn,prb,showpacs]{revtex4}
\usepackage{graphicx}
\usepackage{dcolumn}
\usepackage{bm}
\usepackage{amsmath}

\begin{document}

\preprint{}
\title{ Spin and valley transports in junctions of Dirac fermions}
\author{Takehito Yokoyama}
\affiliation{Department of Physics, Tokyo Institute of Technology, Tokyo 152-8551,
Japan 
}
\date{\today}

\begin{abstract}
We study spin and valley transports in junctions composed of silicene and topological crystalline insulators. 
We consider normal/magnetic/normal Dirac metal junctions where a gate electrode is attached to the magnetic region. 
In normal/antiferromagnetic/normal silicene junction, we show that the current through this junction is valley and spin polarized due to the coupling between valley and spin degrees of freedom, and the valley and spin polarizations can be tuned by local application of a gate voltage. In particular, we find a fully valley and spin polarized current by applying the electric field. 
In normal/ferromagnetic/normal topological crystalline insulator junction with a strain induced in the ferromagnetic segment, we investigate valley resolved conductances and clarify how the valley polarization stemming from the strain and exchange field appears in this junction. It is found that changing the direction of the magnetization and the potential in the ferromagnetic region, one can control the dominant valley contribution out of four valley degrees of freedom.
 We also review spin transport in  normal/ferromagnetic/normal graphene junctions, and spin and valley transports in normal/ferromagnetic/normal silicene junctions for comparison.

\end{abstract}

\maketitle


\section{Introduction}


There has been a great interest in graphene due to its rich potential from fundamental and applied physics point of view. \cite{Ando,Katsnelson,Castro}
Graphene is composed of carbon atoms on a two-dimensional honeycomb lattice. Consequently, electrons in graphene obey the massless Dirac equation.
The recent experimetal progress of fabrication of single graphene sheets has triggered tremendous interests from the scientific community.\cite{novoselov,zhang,novoselov_nature}
Up to now, many intriguing aspects of graphene have been revealed, such as half integer and unconventional quantum Hall effect\cite{zhang,Novoselov2,Yang}, minimum conductivity\cite{novoselov_nature}, and the Klein tunneling\cite{katsnelson,Cheianov,Cayssol,Huard,Williams,Ozyilmaz,Stander,Young,Young2}.
Graphene is also a suitable material for applications: it exhibits gate-voltage-controlled carrier conduction, high field-effect mobilities and a small spin-orbit interaction.\cite{Kane,Hernando}  Therefore, graphene offers a good testing ground for observing spintronics effects. \cite{Son,Kan,Yazyev,Haugen,Tombros,Ohishi,Cho,Yokoyama,Linder2,Yokoyama2}
It has been shown that zigzag edge graphene nanoribbon becomes half-metallic by an external transverse electric field due to the different chemical potential shift at the edges.\cite{Son,Kan,Yazyev} This indicates the high controllability of ferromagnetism in graphene and hence paves the way for spintronics application of graphene. 
In graphene covered by ferromagnet, spin transport controlled by a gate electrode has been predicted. \cite{Haugen,Yokoyama,Yokoyama2}
Also, there are some attempts to use pseudospin (sublattice) degrees of freedom in graphene in order to obtain new functionalities. \cite{Jose,Xia,Majidi}

The goal of valleytronics is to manipulate valley degrees of freedom by electric means and vice versa. This field has developed in graphene\cite{Rycerz,Xiao,Akhmerov}, because graphene has two inequivalent Dirac cones at $K$ and $K'$ points, which can be considered as valley degree of freedom.
In graphene nanoribbons with a zigzag edge, 
valley filter and valley valve effect have been predicted.\cite{Rycerz,Akhmerov} These stem from intervalley scatterings by a potential step and are thus controllable by local application of a gate voltage.

Silicene is a monolayer of silicon atoms on a two dimensional honeycomb lattice: the silicon analog of graphene.\cite{Takeda}  Recently, it has been reported that this material has been synthesized.\cite{Lalmi,Padova,Padova2,Vogt,Lin,Fleurence} Although silicene is composd of silicon atoms on honeycomb lattice and hence electrons in silicene obey the Dirac equation around the $K$ and $K'$ points at low energy\cite{Liu,Liu2}, there are a few important differences from graphene: (i) the honeycomb lattice is buckled. Hence, the mass of the Dirac electrons in silicene can be manipulated by external electric field.\cite{Ezawa,Ezawa2} The discovery of this property has triggered many intriguing predictions.  It has been predicted that there occurs a topological phase transition between topologically trivial and topological insulators by applying electric field. \cite{Ezawa,Ezawa2}
(ii) silicene has a large spin-orbit coupling compared to graphene which couples spin and valley degrees of freedom. Therefore, one may expect interesting spin and valley coupled physics in silicene.

Topological crystalline insulators are new states of matter defined by a topological invariant constructed by crystal symmetries.\cite{Fu,Hsieh,Slager}
Topological crystalline insulators possess even number of gapless surface states on crystal faces that preserve the underlying symmetry. These gapless surface states are topologically protected: they are robust against perturbations as long as the underlying symmetry is preserved.
The (001) surface states composed of four Dirac cones in Pb$_x$Sn$_{1-x}$(Te, Se), the first topological crystalline insulator material, have been predicted\cite{Hsieh} and observed in angle-resolved photoemission spectroscopy experiments\cite{Tanaka,Xu,Dziawa,Wojek,Tanaka2}.
Recently, measurement of surface transport in epitaxial SnTe thin films has been also reported. \cite{Taskin}
Since these materials have four Dirac cones in contrast to honeycomb systems, 
 topological crystalline insulators have a potential to be placed ahead of graphene for valleytronics applications.\cite{Ezawa4}

In this paper, 
we first review spin transport in normal/ferromagnetic/normal graphene junctions, and spin and valley transports in normal/ferromagnetic/normal silicene junctions.
Then, we study spin and valley transports in junctions composed of silicene and topological crystalline insulators. 
We consider normal/magnetic/normal Dirac metal junctions where a gate electrode is attached to the magnetic region. 
In normal/antiferromagnetic/normal silicene junction, we show that the current through this junction is valley and spin polarized due to the coupling between valley and spin degrees of freedom, and the valley and spin polarizations can be tuned by local application of a gate voltage. In particular, we find a fully valley and spin polarized current by applying the electric field. 
In normal/ferromagnetic/normal topological crystalline insulator junction with a strain induced in the ferromagnetic segment, we investigate valley resolved conductances and clarify how the valley polarization stemming from the strain and exchange field appears in this junction. It is found that changing the direction of the magnetization and the potential in the ferromagnetic region, one can control the dominant valley contribution out of four valley degrees of freedom.

\section{Graphene}

\begin{figure}[tbp]
\begin{center}
\scalebox{0.8}{
\includegraphics[width=12.0cm,clip]{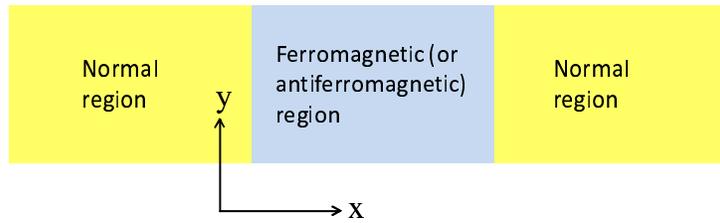}
}
\end{center}
\caption{ Schematic picture of the model of junctions of Dirac fermions. }
\label{fig1}
\end{figure}


Here, we review spin transport in normal/ferromagnetic/normal graphene junctions, following Ref.\cite{Yokoyama}  .

The electrons in graphene obey a massless Dirac equation given by 
\begin{equation}
H_\pm = v_F (\sigma_x k_x + \eta \sigma_y k_y)
\end{equation}
with Pauli matrices $\sigma_x$ and $\sigma_y$ which operate on the sublattice space of the honeycomb lattice.  
 The $\eta=\pm$ sign corresponds to the two valleys of $K$ and $K'$ points in the Brillouin zone. Also, there is a valley degeneracy. Hence, one can consider one of the two valleys ($H_\pm$). \cite{morpurgo2}
The linear dispersion relation is valid for Fermi levels up to 1 eV, \cite{wallace} where the electrons in graphene behave like Weyl fermions in the low-energy regime.

We consider a two dimensional normal/ferromagnetic/normal graphene junction where a gate electrode is attached to the ferromagnetic region. This junction may be realized by putting a ferromagnetic insulator on top of graphene or doping magnetic atoms into graphene. 
  See Figure \ref{fig1} for the schematic of the model.  We assume that the interfaces are parallel to the $y$-axis and located at $x=0$ and $x=L$. Due to the valley degeneracy, we consider the Hamiltionian $H_+$ with $H_+ = v_F (\sigma_x k_x + \sigma_y k_y) - V(x)$,  $V(x)=E_F$ in the normal graphenes and $V(x)= E_F  + U \pm  H$ in the ferromagnetic graphene. Here, $E_F=v_F k_F$ is the Fermi energy, $U$ is the potential shift controllable by the gate voltage, and $H$ is the exchange field. Here, $\pm$ signs correspond to majority and minority spins. 
The wavefunctions in each regions can be written as 
\begin{eqnarray}
 \psi_1 = \left( {\begin{array}{*{20}c}
   1  \\
   {e^{i\theta } }  \\
\end{array}} \right)e^{ip\cos \theta x +ip_y y} 
 + a_\pm \left( {\begin{array}{*{20}c}
   1  \\
   { - e^{ - i\theta } }  \\
\end{array}} \right)e^{ - ip\cos \theta x +ip_y y}  ,
\end{eqnarray}
\begin{eqnarray}
 \psi_2 = b_\pm \left( {\begin{array}{*{20}c}
   1  \\
   {e^{i\theta '} }  \\
\end{array}} \right)e^{ip'_\pm \cos \theta 'x +ip_y y} 
+ c_\pm \left( {\begin{array}{*{20}c}
   1  \\
   { - e^{ - i\theta '} }  \\
\end{array}} \right)e^{ - ip'_\pm \cos \theta 'x +ip_y y}  ,
\end{eqnarray}
\begin{eqnarray}
 \psi_3 = d_\pm \left( {\begin{array}{*{20}c}
   1  \\
   {e^{i\theta } }  \\
\end{array}} \right)e^{ip\cos \theta x +ip_y y} 
\end{eqnarray}
with angles of incidence $\theta$ and $\theta'$, $p = (E + E_F )/v_F$ and $ p'_\pm  = (E + E_F  + U \pm  H)/v_F$. 
Here, $\psi_1$ and $\psi_3$ denote wavefunctions in the left and right normal graphenes, respectively, while $\psi_2$ is a wavefunction in the ferromagnetic graphene. 
Due to the translational invariance in the $y$-direction, the momentum parallel to the $y$-axis is conserved: 
$p_y= p\sin \theta  = p'\sin \theta '$.

By matching the wave functions at the interfaces, we obtain the coefficients in the above wavefunctions in Eqs.(2-4). 
Note that these conditions lead to the current conservation at the interfaces because they are reduced to $\hat v_x \psi_1=\hat v_x \psi_2$ at $x = 0$ and $\hat v_x \psi_2=\hat v_x \psi_3$ at $x = L$ where $\hat v_x $ is the velocity operator given by $\hat v_x  = \partial H_ +  /\partial k_x  = v_F\sigma _x$.

The transmission coefficient is represented as 
\begin{eqnarray}
d_\pm = \frac{{\cos \theta \cos \theta '{\mathop{\rm e}\nolimits} ^{ - ipL\cos \theta } }}{{\cos(p'_\pm L\cos \theta ')\cos \theta \cos \theta ' - i\sin (p'_\pm L\cos \theta ')(1 - \sin \theta \sin \theta ')}}.
\end{eqnarray}

Thus, the dimensionless spin-resolved conductances $G_{\uparrow,\downarrow}$ are obtained as 
\begin{eqnarray}
G_{\uparrow,\downarrow}   = \frac{1}{2}\int_{ - \pi /2}^{\pi /2} {d\theta \cos \theta T_{\uparrow,\downarrow}  (\theta )} 
\end{eqnarray}
with $T_{\uparrow,\downarrow} (\theta ) = \left| {d_\pm(\theta )} \right|^2$. 
Finally, the spin conductance $G_s$ is defined as 
$G_s  = G_ \uparrow   - G_ \downarrow$. Below, we focus on the conductances at zero voltage, setting $E=0$. 

First, we will explain the underlying mechanism of spin manipulation by the gate voltage. 
In the limit of $\left| U \pm H \right| \gg E_F$, we have $\theta' \to 0$, and therefore, the transmission coefficient becomes
\begin{eqnarray}
d_\pm \to \frac{{\cos \theta {\mathop{\rm e}\nolimits} ^{ - ipL\cos \theta } }}{{\cos\chi_\pm  \cos \theta  - i\sin \chi_\pm  }}
\end{eqnarray}
with $\chi _\pm   = \chi  \pm \chi _H, \chi  = UL/v_F,$ and $\chi _H  =HL/v_F$. 
The transmission probability is thus given by \cite{katsnelson}
\begin{eqnarray}
T_{ \uparrow , \downarrow } (\theta ) \to \frac{{\cos ^2 \theta }}{{1 - \sin ^2 \theta \cos^2 \chi _\pm }}.
\end{eqnarray}
From Eq.(8), we find the $\pi$-periodicity with respect to $\chi_\pm$ or $\chi$. \cite{Haugen,katsnelson,linder,sengupta}
 It is also seen that $G_{\uparrow,\downarrow} $ has a maximum (minimum) value of 1 (2/3) at $\chi_\pm=0$ $(\pi/2)$. 
The phase difference between $G_{\uparrow}$ and $G_{\downarrow} $ is given by $\chi _+ - \chi_- = 2\chi _H  = 2HL/v_F$. If the phase difference is equal to the half period $\pi/2$ (i.e., $H/E_F= \pi /4 k_F L$), one can expect a large spin current which oscillates with $\chi$, i.e., the gate voltage, because when one of $G_{\uparrow}$ and $G_{\downarrow} $ has a maximum at a certain $\chi$, the other has a minimum at the same $\chi$. As a result, the value of $G_s$ oscillates between $-1/3$ and $1/3$. Notice that the electrical conductance $G_{\uparrow}+G_{\downarrow} $ in the junctions is always positive and hence spin current reversal in our model is not accompanied with the charge current reversal.

\begin{figure}[htb]
\begin{center}
\scalebox{0.4}{
\includegraphics[width=25.0cm,clip]{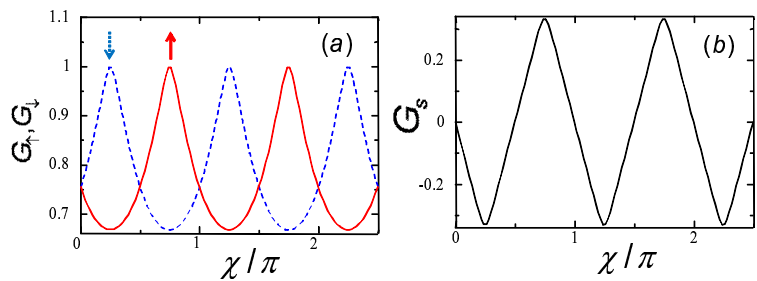}}
\end{center}
\caption{ Plots of conductances as a function of $\chi$ tunable by  the gate voltage, in the limit of $U \to \infty$. (a) spin resolved conductances where the phases of 
$G_ \uparrow$ (solid line) and $G_ \downarrow$ (dotted line) are shifted by half period, $\chi _\uparrow - \chi _\downarrow= \pi/2$. (b) Spin conductance $G_s$ which oscillates with the period $\pi$ with respect to $\chi$ but is never damped. } \label{f2}
\end{figure}

In Fig. \ref{f2}, we show the results in this limiting case. 
 Figure \ref{f2} (a) depicts spin resolved conductances as a function of $\chi$. Here, the phases of $G_ \uparrow$ and $G_ \downarrow$ are shifted by half period, $\chi _+ - \chi _-= \pi/2$. 
As shown in  Fig. \ref{f2} (b), we obtain a finite spin current. Interestingly,  the spin conductance oscillates with the period $\pi$ with respect to $\chi$. This indicates that one can reverse the spin current by changing the gate voltage. Here, we have focused on the limiting case. For more general cases, see Ref.\cite{Yokoyama}.

\section{Silicene}
Spin and valley transports in normal/ferromagnetic/normal silicene junction have been studied in Ref.\cite{Yokoyama3}. Here, we review  spin and valley transports in this junction and investigate them in a normal/antiferromagnetic/normal silicene junction as shown in Fig. \ref{fig1}.

\subsection{Formulation}
The Hamiltonian of the (anti)ferromagnetic silicene is given by \cite{Liu,Liu2,Ezawa,Ezawa2}
\begin{eqnarray}
H = \hbar v_F (k_x \tau _x  - \eta k_y \tau _y ) - \Delta _{\eta \sigma } \tau _z  - \sigma h
\end{eqnarray}
with $\Delta _{\eta \sigma }  = \eta \sigma \Delta _{so}  - \Delta _z  + \sigma h_s $. $\tau$ is the Pauli matrix in sublattice pseudospin space. $\Delta _{so}$ denotes the spin-orbit coupling. $\Delta _z$ is the onsite potential difference between $A$ and $B$ sublattices, which can be manipulated by an electric field applied perpendicular to the plane. $h (h_s)$ is the ferromagnetic (antiferromagnetic or staggered) exchange field in the ferromagnetic (antiferromagnetic) region. $\eta  =  \pm 1$ corresponds to the $K$ and $K'$ points. $\sigma  =  \pm 1$ denotes the spin indices. 
The large value of $\Delta _{so}= 3.9$ meV in silicene \cite{Liu2} leads to a coupling between valley and spin degrees of freedom, which is a clear distinction from graphene.
In the normal regions, we set $\Delta _z=h=h_s=0$. 
Thus, the gate electrode is attached to the magnetic segment. The eigenvalues of the Hamiltonian in the normal and magnetic silicene are given by 
\begin{eqnarray}
E =  \pm \sqrt {(\hbar v_F k)^2  + (\Delta _N )^2 } 
 =  \pm \sqrt {(\hbar v_F k')^2  + (\Delta _F )^2 }  - \sigma h
\end{eqnarray}
with $\Delta _N  = \eta \sigma \Delta _{so}$ and $\Delta _F  = \eta \sigma \Delta _{so}  - \Delta _z + \sigma h_s$. $k$ and $k'$ are momenta in the normal and the magnetic regions, respectively. 
Let $x$-axis perpendicular to the interface and assume the translational invariance along the $y$-axis.  The interfaces between the normal and the magnetic silicene are located at $x=0$ and $x=L$ where $L$ is the length of the magnetic silicene. Then, the wavefunctions for valley $\eta$ and spin $\sigma$ in each region can be written as 
\begin{widetext}
\begin{eqnarray}
 \psi (x < 0) = \frac{1}{{\sqrt {2EE_N } }}e^{ik_x x} \left( {\begin{array}{*{20}c}
   {\hbar v_F k_ +  }  \\
   {E_N }  \\
\end{array}} \right) + \frac{{r_{\eta ,\sigma } }}{{\sqrt {2EE_N } }}e^{ - ik_x x} \left( {\begin{array}{*{20}c}
   { - \hbar v_F k_ -  }  \\
   {E_N }  \\
\end{array}} \right), \\ 
 \psi (0 < x < L) = a_{\eta ,\sigma } e^{ik'_x x} \left( {\begin{array}{*{20}c}
   {\hbar v_F k'_ +  }  \\
   {E_F }  \\
\end{array}} \right) + b_{\eta ,\sigma } e^{ - ik'_x x} \left( {\begin{array}{*{20}c}
   { - \hbar v_F k'_ -  }  \\
   {E_F }  \\
\end{array}} \right), \\ 
 \psi (L < x) = \frac{{t_{\eta ,\sigma } }}{{\sqrt {2EE_N } }}e^{ik_x x} \left( {\begin{array}{*{20}c}
   {\hbar v_F k_ +  }  \\
   {E_N }  \\
\end{array}} \right)
\end{eqnarray}
\end{widetext}
with $\hbar v_F k'_x  = \sqrt {(E + \sigma h)^2  - (\Delta _F )^2  - (\hbar v_F k_y )^2 } $, $ E_N  = E + \Delta _N ,E_F  = E + \sigma h + \Delta _F$, and $k_ \pm ^{(\prime)}  = k_x^{(\prime)}  \pm i\eta k_y$. 
Here, ${r_{\eta ,\sigma } }$ and ${t_{\eta ,\sigma } }$ are reflection and transmission coefficients, respectively.
By matching the wavefunctions at the interfaces, we obtain the transmission coefficient: 
\begin{eqnarray}
 t_{\eta ,\sigma }  = 4k_x k'_x E_N E_F e^{ - ik_x L} /A,\quad  \\ 
  A = (\alpha ^{ - 1}  - \alpha )k^2 E_F^2  + (\alpha ^{ - 1}  - \alpha )(k')^2 E_N^2 
  + E_N E_F \left[ {k_ +  (\alpha ^{ - 1} k'_ +   + \alpha k'_ -  ) + k_ -  (\alpha ^{ - 1} k'_ -   + \alpha k'_ +  )} \right]  
\end{eqnarray}
with $\alpha  = e^{ik'_x L}$. 

By setting $k_x  = k\cos \phi$ and $k_y  = k\sin \phi$,  we define normalized valley and spin resolved conductance: 
\begin{eqnarray}
G_{\eta \sigma }  = \frac{1}{2}\int_{ - \pi /2}^{\pi /2} {\left| {t_{\eta, \sigma } } \right|^2 \cos \phi d\phi } .
\end{eqnarray}
The valley and spin resolved conductances, $G_{K^{(')}}$ and $G_{ \uparrow ( \downarrow )}$, and valley and spin polarizations, $G_v$ and $G_s$, are defined as follows:
\begin{eqnarray}
 G_{K^{(')}}  = \frac{{G_{{K^{(')}} \uparrow }  + G_{{K^{(')}} \downarrow } }}{2}, \\ 
 G_{ \uparrow ( \downarrow )}  = \frac{{G_{K \uparrow ( \downarrow )}  + G_{K' \uparrow ( \downarrow )} }}{2}, \\ 
G_v  = \frac{{G_K  - G_{K'} }}{{G_K  + G_{K'} }}, \\ 
G_s  = \frac{{G_ \uparrow   - G_ \downarrow  }}{{G_ \uparrow   + G_ \downarrow  }} .
\end{eqnarray}

\subsection{Results}

\begin{figure}[tbp]
\begin{center}
\scalebox{0.8}{
\includegraphics[width=13.50cm,clip]{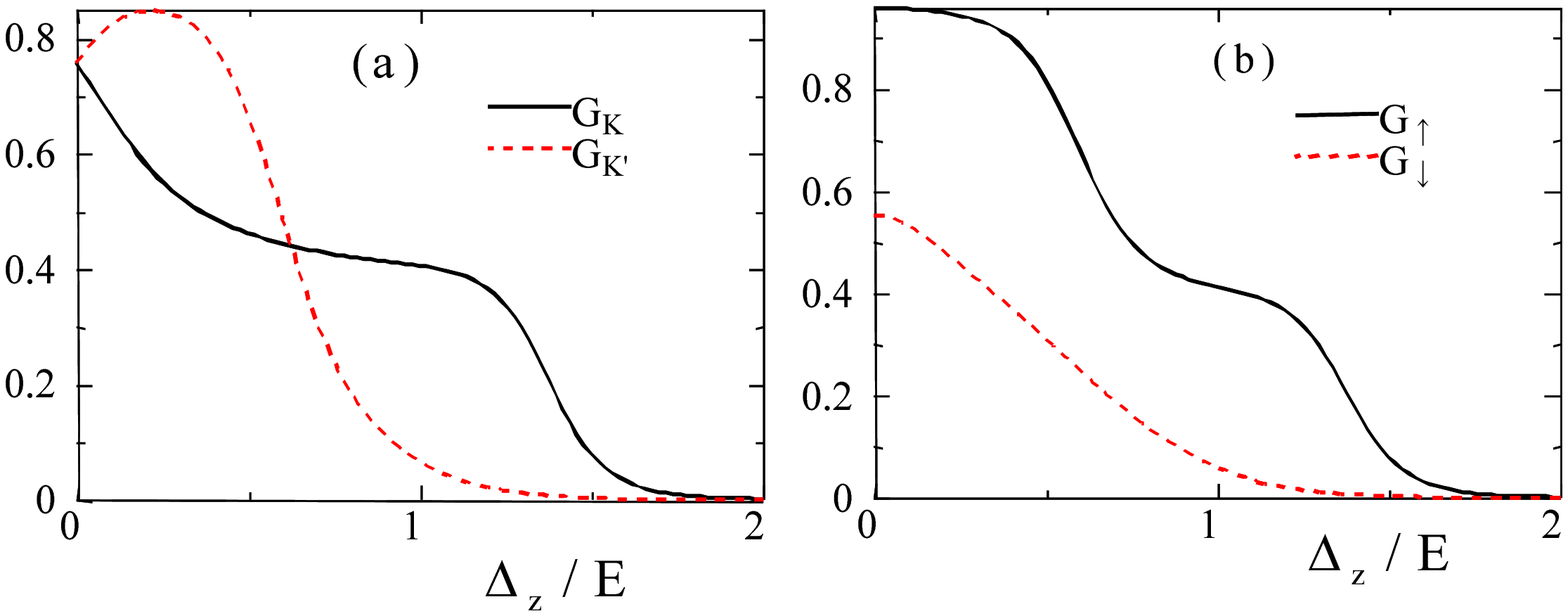}
}
\end{center}
\caption{ (a) Valley resolved conductance $G_{K(K')}$ as a function of $\Delta_z$. (b) Spin resolved conductance $G_{ \uparrow ( \downarrow )}$ as a function of $\Delta_z$. We set $h/E=0.3$ and $h_s=0$.}
\label{fig3}
\end{figure}

\begin{figure}[tbp]
\begin{center}
\scalebox{0.8}{
\includegraphics[width=17.50cm,clip]{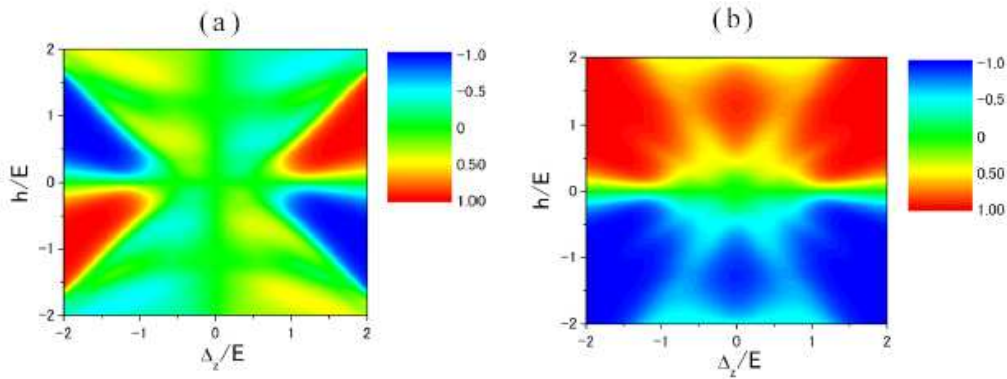}
}
\end{center}
\caption{ (a) Valley polarization $G_v$ as functions of $\Delta_z$ and $h$. (b) Spin polarization $G_s$ as functions of $\Delta_z$ and $h$. We set $h_s=0$.}
\label{fig4}
\end{figure}

In the following, we fix $L$ and $\Delta_{so}$ as $k_F L=3$ and $\Delta_{so}/E=0.5$ where $k_F=E/(\hbar v_F)$.
We consider a finite chemical potential by doping in silicene.

First, let us review the ferromagnetic junctions with $h_s=0$. \cite{Yokoyama3}
Figure \ref{fig3} depicts  (a) valley resolved conductance $G_{K(K')}$ and (b) spin resolved conductance $G_{ \uparrow ( \downarrow )}$ as a function of $\Delta_z$. 
As seen from Figure \ref{fig3} (a), with increasing $\Delta_z$, the current stemming from the $K'$ point strongly decreases.
Then, $G_K$ gives a dominant contribution to the current. We find that $G_{ \uparrow}$ dominates over $G_{\downarrow}$ for large $\Delta_z$ as seen in Fig. \ref{fig3}(b). These behaviors are attributed to the band structures in the ferromagnetic region.\cite{Yokoyama3}
Figure \ref{fig4} illustrates (a) $G_v$ and (b) $G_s$ as functions of $\Delta_z$ and $h$ for $h_s=0$. The valley polarization $G_v$ is odd with respect to $\Delta_z$ and $h$. For large  $\Delta_z$, $G_v$ becomes large as we found in Fig. \ref{fig3} (a). However, for smaller $\Delta_z$, the magnitude of $G_v$ can be still $\sim 0.5$. We find that even the sign of the valley polarization can be changed by varying $\Delta_z$. It is also seen that $G_v$ changes significantly by varying the exchange field $h$. This indicates that the valley polarization can be manipulated magnetically. 
The spin polarization $G_s$ is odd in $h$ but even in $\Delta_z$. For large $h$, $G_s$ becomes large as expected. Even for small $h$, $G_s$ can be large for large $\Delta_z$.
From Figure \ref{fig4}, it is also found that fully valley and spin polarized currents are realized for large $\Delta_z$ but relatively high polarizations ($\ge  0.5$) can be realized in a wide parameter regime.

The condition to realize fully valley polarized transport can be obtained as follows. For simplicity, let us focus on the regime with $\Delta _z>0$ and $h>0$.
To locate the Fermi level $E (> \Delta_{so})$ within the band gap at the $K'$ point ($\eta  =  - 1$), $- \left| {\sigma \Delta _{so}  + \Delta _z } \right| - \sigma h < \Delta _{so}  < \left| {\sigma \Delta _{so}  + \Delta _z } \right| - \sigma h$ should be satisfied. Therefore, we obtain the condition necessary for the fully valley polarized transport as
\begin{eqnarray}
\Delta _z  > \max (h, \Delta _{so} , 2\Delta _{so}  - h) .
\end{eqnarray}

\begin{figure}[tbp]
\begin{center}
\scalebox{0.8}{
\includegraphics[width=13.50cm,clip]{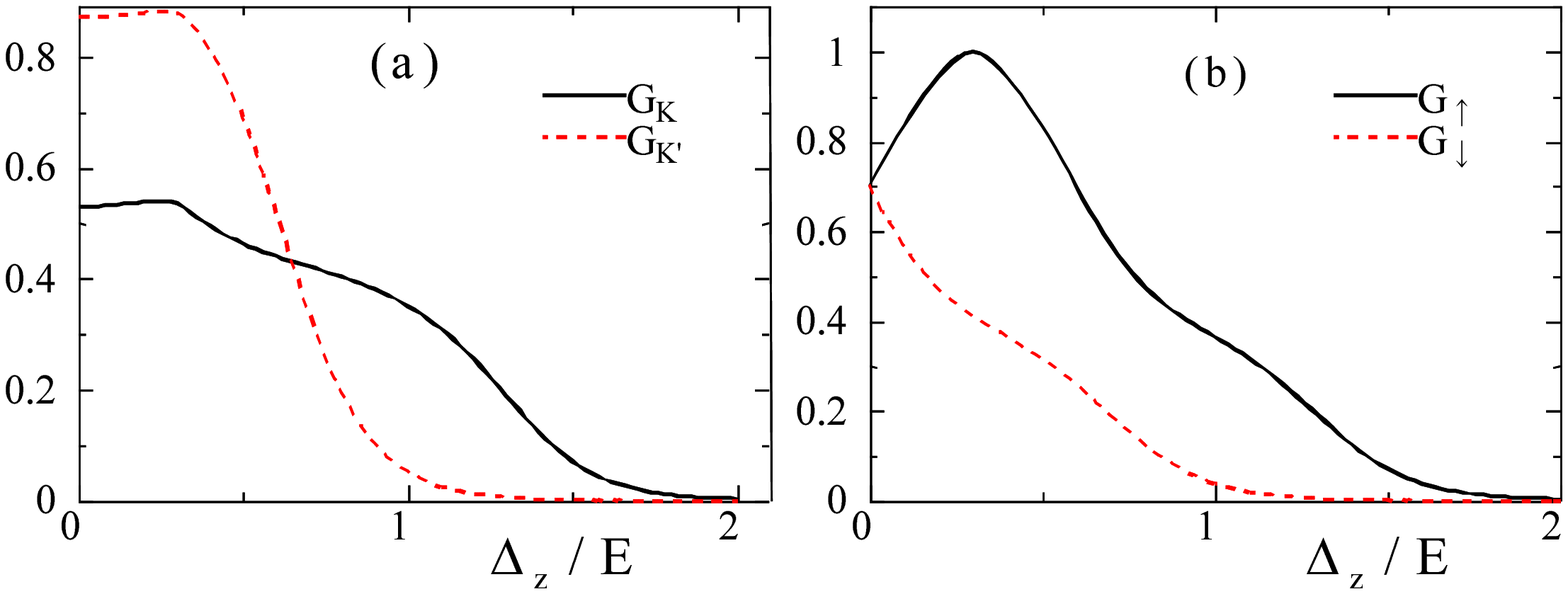}
}
\end{center}
\caption{  (a) Valley resolved conductance $G_{K(K')}$ as a function of $\Delta_z$. (b) Spin resolved conductance $G_{ \uparrow ( \downarrow )}$ as a function of $\Delta_z$. We set $h=0$ and $h_s/E=0.3$.}
\label{fig5}
\end{figure}

\begin{figure}[tbp]
\begin{center}
\scalebox{0.8}{
\includegraphics[width=17.50cm,clip]{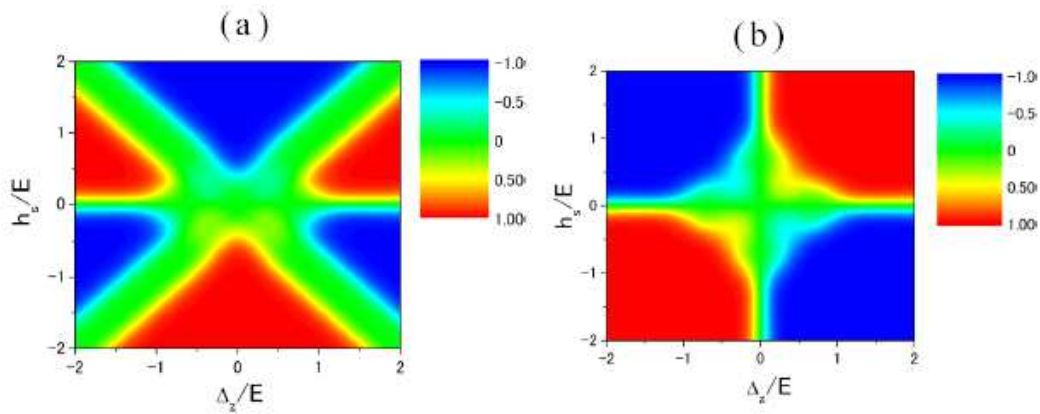}
}
\end{center}
\caption{ (a) Valley polarization $G_v$ as functions of $\Delta_z$ and $h_s$. (b) Spin polarization $G_s$ as functions of $\Delta_z$ and $h_s$. We set $h=0$. }
\label{fig6}
\end{figure}

Next, consider the antiferromagnetic junctions with $h=0$. 
Figure \ref{fig5} depicts  (a) valley resolved conductance $G_{K(K')}$ and (b) spin resolved conductance $G_{ \uparrow ( \downarrow )}$ as a function of $\Delta_z$. 
As seen from Figure \ref{fig5} (a), with increasing $\Delta_z$, the current coming from the $K'$ point strongly decreases.
Then, $G_K$ gives a dominant contribution to the current. We also find that $G_{ \uparrow}$ dominates over $G_{\downarrow}$ for large $\Delta_z$ as seen from Fig. \ref{fig5}(b).  These behaviors are again attributed to the band structures in the antiferromagnetic region.

Figure \ref{fig6} shows (a) $G_v$ and (b) $G_s$ as functions of $\Delta_z$ and $h_s$ for $h=0$. The valley polarization $G_v$ is an even function of $\Delta_z$ but an odd function of $h_s$. For large  $\Delta_z$, $G_v$ becomes large, which is consistent with Fig. \ref{fig5} (a).  We find that the sign of the valley polarization can be changed by varying $\Delta_z$. 
In constrast to the ferromagnetic case with finite $h$, the $G_v$ can be large for $\Delta_z=0$ but with finite $h_s$. 
It is also seen that $G_v$ changes significantly by varying the exchange field $h_s$. This again indicates that the valley polarization can be controlled magnetically. 
The spin polarization $G_s$ is odd in $h_s$ and $\Delta_z$. 
Thus, the $G_s$ becomes zero at $\Delta_z=0$. This can be also understood from the fact that the bands are spin degenerate for $\Delta_z=0$. 
Even for small $h_s$, $G_s$ can be large for large $\Delta_z$.
From this figure, it is found that fully valley and spin polarized currents are realized for large $\Delta_z$ and $h_s$ regime. Interestingly, we find some parameter regions where $G_v=0$ but $G_s=\pm1$ or $G_v=\pm1$ but $G_s=0$. Namely, by changing the tunable parameter $\Delta_z$, one can realize transitions from a fully valley  polarized state without spin polarization to a fully spin polarized state without valley polariztion. 

The conditions to realize fully valley or spin polarized transports are obtained as follows. Let us focus on the regime with $\Delta_z, h_s>0$. 
The gap for valley $\eta$ and spin $\sigma$ is given by $\left| {\Delta _F } \right| = \left| {\eta \sigma \Delta _{so}  - \Delta _z  + \sigma h_s } \right|$.
 To realize the fully valley polarized transport $G_v=1$, the gaps at the $K'$ point should be larger than the Fermi energy: $\left| { - \sigma \Delta _{so}  - \Delta _z  + \sigma h_s } \right| > E
$. Thus, we obtain 
\begin{eqnarray}
- \Delta _z + E +\Delta _{so}< h_s <\Delta _z - E +\Delta _{so}. 
\end{eqnarray}
For $\Delta _{so} /E=0.5$, this reduces to $ - \Delta _z /E + 1.5 < h_s/E <\Delta _z /E -0.5$, which is consistent with Fig. \ref{fig6} (a). 
Similarly, to obtain $G_v=-1$, we require $\left| { \sigma \Delta _{so}  - \Delta _z  + \sigma h_s } \right| > E$, leading to 
\begin{eqnarray}
\Delta _z + E - \Delta _{so}< h_s .
\end{eqnarray}
For $\Delta _{so} /E=0.5$, this reduces to $ \Delta _z / E +0.5< h_s/E$, which is consistent with Fig. \ref{fig6} (a). 
To obtain the fully spin polarized transport, $G_p=1$, the gaps for spin down states are required to be larger than the Fermi energy: $\left| { \Delta _{so}  + \Delta _z  + h_s } \right|, \left| { \Delta _{so}  -\Delta _z  -h_s } \right| > E$.
Thus, we obtain 
\begin{eqnarray}
- \Delta _z + E +\Delta _{so}< h_s.
\end{eqnarray}
For $\Delta _{so} /E=0.5$, this reduces to $ \Delta _z / E +1.5< h_s/E$, consistent with Fig. \ref{fig6} (b). 
Also, note that when $\Delta _z = h_s$ and $2 \Delta _z > E+\Delta _{so}$ are satisfied, the gaps for spin up states at the $K$ and $K'$ points coincide and the gaps for spin down states are larger than the Fermi energy. Thus, we have $G_v=0$ and $G_p=1$ in this case, as seen in Fig. \ref{fig6}.

For a ferromagnetic silicene with $k_F L$ = 1 and $E=$10 meV, since $v_F \sim 5 \times 10^5$m/s, we have $L \sim$ 10 nm. 
For $\Delta_z \sim E$, the electric field applied perpendicular to the plane is  estimated as 34 meV/\AA $ $
 since the distance between the $A$ and $B$ sublattice planes is 0.46 \AA. 
Here, we have assumed the zero temperature limit. This assumption is justified for tempereture regime lower than $\Delta_{so}$, $\Delta_z$, $h$ and $h_s$. 

Recently, based on first-principles calculations, stability and electronic structures of silicene on Ag(111) surfaces have been investigated.\cite{Guo,Wang} It is found that Dirac electrons are absent near Fermi level in all the stable structures due to buckling of the Si monolayer and mixing between Si and Ag orbitals. It is also proposed that either BN substrate or hydrogen-processed Si surface is a good candidate to preserve Dirac electrons in silicene. \cite{Guo}

A ferromagnetic exchange field  could be induced in silicene by the magnetic proximity effect with a magnetic insulator EuO as proposed for graphene, which could be of the order of 1 meV.\cite{Haugen} 
Exchange fields on A and B sublattices can be induced by sandwiching silicene by two (different) ferromagnets or attaching a honeycomb-lattice antiferromagnet such as antiferromagnetic manganese chalcogenophosphates (MnPX$_3$, X = S, Se) in monolayer form.\cite{Ezawa3,Liang,Li}

\section{Topological crystalline insulator}

\subsection{Formulation}
Consider normal/ferromagnetic/normal topological crystalline insulator junctions with flat interfaces at $x=0$ and $x=L$ (See Fig. \ref{fig1}).
We here study transports on the (001) surface of the topological crystalline insulator.
The topological crystalline insulator has four Dirac cones with the same chirality at $\Lambda_X $, $\Lambda'_{X}$, $\Lambda_Y $, and $\Lambda'_{Y}$ points in the (001) surface. \cite{Liu3,Fang,Liu4,Wang2,Fan}
The effective Hamiltonian of the topological crystalline insulator around the $\Lambda _X $ point is given by \cite{Liu3,Fang,Liu4,Wang2,Fan}
\begin{eqnarray}
H_X  = v_1 \tilde k_x \sigma _y  - v_2 \tilde k_y \sigma _x  + \tilde m\sigma _z + U
\end{eqnarray}
where typically $v_1=1.3$eV\AA, $v_2=0.84$eV\AA, $\sigma$ is the Pauli matrix in spin space, and
\begin{eqnarray}
\tilde k_x  = k_x  + \frac{1}{{v_1 }}\left( {\lambda _{11} \varepsilon _{11}  + \lambda _{22} \varepsilon _{22}  + \lambda _{33} \varepsilon _{33}  + h_y } \right),\;\tilde k_y  = k_y  - \frac{1}{{v_2 }}\left( {\lambda _{12} \varepsilon _{12}  + h_x } \right),\\  \tilde m = \frac{{n'}}{{\sqrt {n^2  + (n')^2 } }}\left( {\lambda _{23} \varepsilon _{23}  + h_z } \right) \cong 0.35\left( {\lambda _{23} \varepsilon _{23}  + h_z } \right). 
\end{eqnarray}
Here, $U$ is the potential, $\varepsilon _{ij}$ and $\lambda _{ij}$ $(i, j=1, 2, 3)$ are the strain tensor and electron-phonon couplings of the topological crystalline insulator, respectively.
Strain may be induced by substituting Se for Sn,\cite{Okada} or by attaching a piezoelectric material such as BaTiO$_3$. \cite{Fan}
 $n=70$meV and $n'=26$meV describe the intervalley scattering. \cite{Liu3,Fang,Liu4,Wang2,Ezawa4}
$h_i$  $(i=x, y, z)$ represents the induced exchange field in the ferromagnetic region given by
\begin{eqnarray}
(h_x ,h_y ,h_z ) = h(\sin \theta \cos \varphi ,\sin \theta \sin \varphi ,\cos \theta ). 
\end{eqnarray}
We set $\varepsilon _{ij}=h=U=0$ in the normal regions and consider a scattering problem through a barrier region induced by the ferromagnetism and strain. 

The wavefunctions in each regions can be written as 
\begin{eqnarray}
 \psi (x \le 0) = \frac{1}{{\sqrt 2 E}}e^{ik_x x} \left( {\begin{array}{*{20}c}
   { - iv_1 k_x  - v_2 k_y }  \\
   E  \\
\end{array}} \right) + \frac{r}{{\sqrt 2 E}}e^{ - ik_x x} \left( {\begin{array}{*{20}c}
   {iv_1 k_x  - v_2 k_y }  \\
   E  \\
\end{array}} \right), \\ 
 \psi (0 < x < L) = \frac{a}{{\sqrt {2E'(E' - \tilde m)} }}e^{ik'_x x} \left( {\begin{array}{*{20}c}
   { - iv_1 k'_x  - v_2 \tilde k_y }  \\
   {E' - \tilde m}  \\
\end{array}} \right) + \frac{b}{{\sqrt {2E'(E' - \tilde m)} }}e^{ - ik'_x x} \left( {\begin{array}{*{20}c}
   {iv_1 k'_x  - v_2 \tilde k_y }  \\
   {E' - \tilde m}  \\
\end{array}} \right), \\ 
 \psi (x \ge L) = \frac{t}{{\sqrt 2 E}}e^{ik_x x} \left( {\begin{array}{*{20}c}
   { - iv_1 k_x  - v_2 k_y }  \\
   E  \\
\end{array}} \right) .
\end{eqnarray}
Here, $r$ and $t$ are the reflection and transmission coefficients, respectively.
We set $E' = E - U$ and assume that the Fermi energy is positive, $E>0$. The dispersion relations are then given by $E = \sqrt {(v_1 k_x )^2  + (v_2 k_y )^2 }  =  \pm \sqrt {(v_1 k'_x )^2  + (v_2 \tilde k_y )^2  + \tilde m^2 }  + U$. Note that due to the translational symmetry in the $y$-direction, the momentum parallel to the $y$-axis is conserved, while the momentum parallel to the $x$-axis is not conserved.

By matching the wavefunctions at the interfaces,
\begin{eqnarray}
\psi (+0)=\psi (-0), \; \psi (L+0)=\psi (L-0),
\end{eqnarray}
we obtain the transmission coefficient: 
\begin{eqnarray}
t = \frac{{4pv_1 k'_x Ee^{ - ik_x L} \cos \phi }}{{e^{ - ik'_x L} \left( {iv_1 k'_x  + v_2 \tilde k_y  + ipe^{i\phi } } \right)\left( {iv_1 k'_x  - v_2 \tilde k_y  + ipe^{ - i\phi } } \right) + e^{ik'_x L} \left( { - iv_1 k'_x  + v_2 \tilde k_y  + ipe^{i\phi } } \right)\left( {iv_1 k'_x  + v_2 \tilde k_y  - ipe^{ - i\phi } } \right)}}.
\end{eqnarray}
Here, $p = 1 - (U + \tilde m)/E$, and we set $v_1 k_x  = E\cos \phi$ and $v_2 k_y  = E\sin \phi$. 
The normalized conductance stemming from the $\Lambda _X$ point is calculated as
\begin{eqnarray}
G_{X}  = \frac{1}{2}\int_{ - \frac{\pi }{2}}^{\frac{\pi }{2}} {\left| t \right|^2 \cos \phi d\phi } .
\end{eqnarray}
The conductance coming from the $\Lambda'_X$ point $G_{X'}$ can be obtaind by the substitution $\lambda _{23} \varepsilon _{23}  \to  - \lambda _{23} \varepsilon _{23}$ in the above result.\cite{Liu3,Fang,Liu4,Wang2,Fan}

The effective Hamiltonian around the $\Lambda _Y$ point is given by 
\begin{eqnarray}
H_Y  = v_2 \tilde k_x \sigma _y  - v_1 \tilde k_y \sigma _x  + \tilde m\sigma _z+U 
\end{eqnarray}
where 
\begin{eqnarray}
\tilde k_x  = k_x  + \frac{1}{{v_2 }}\left( {\lambda _{11} \varepsilon _{22}  + \lambda _{22} \varepsilon _{11}  + \lambda _{33} \varepsilon _{33}  + h_y } \right),\;\tilde k_y  = k_y  + \frac{1}{{v_1 }}\left( {\lambda _{12} \varepsilon _{12}  - h_x } \right),\;\tilde m = \frac{{n'}}{{\sqrt {n^2  + (n')^2 } }}\left( { - \lambda _{13} \varepsilon _{13}  + h_z } \right).
\end{eqnarray}
The effective Hamiltonian around the $\Lambda' _Y$ point is given by the replacement $\lambda _{13} \varepsilon _{13}  \to  - \lambda _{13} \varepsilon _{13}$ in $H_Y$.\cite{Liu3,Fang,Liu4,Wang2,Fan}
The conductances originating from the $\Lambda_Y$ and $\Lambda' _Y$ points, $G_{Y}$ and $G_{Y'}$, can be obtained in a way similar to that from the $\Lambda_X$ point (by replacement of corresponding parameters).
Note that $G_{Y}$ is given by
\begin{eqnarray}
G_Y  = \frac{{v_2 }}{{2v_1 }}\int_{ - \frac{\pi }{2}}^{\frac{\pi }{2}} {\left| t \right|^2 \cos \phi d\phi } .
\end{eqnarray}
The factor of $\frac{{v_2 }}{{v_1 }}$ is included in this expression because the velocity operator for $H_Y$ is given by
\begin{eqnarray}
\hat v_x  = \frac{{\partial H_Y }}{{\partial k_x }} = v_2 \sigma _y .
\end{eqnarray}

Finally, the total conductance $G$ is defined as 
\begin{eqnarray}
G=G_X+G_{X'}+G_{Y}+G_{Y'} .
\end{eqnarray}

\subsection{Results}

\begin{figure}[tbp]
\begin{center}
\scalebox{0.8}{
\includegraphics[width=17.50cm,clip]{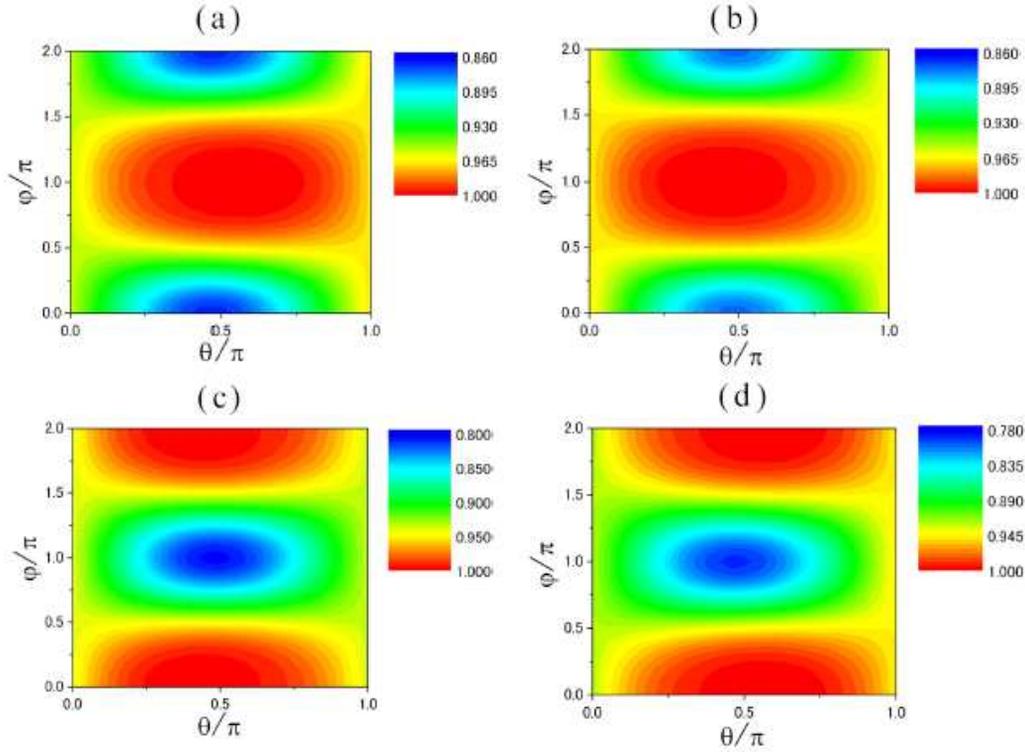}
}
\end{center}
\caption{  Valley resolved conductances (a) $G_{X}$, (b) $G_{X'}$, (c) $G_{Y}$, and (d) $G_{Y'}$ as a function of the direction of the exchange field $\theta$ and $\varphi$ for $U=0$. A valley filtering effect is seen.}
\label{fig7}
\end{figure}

\begin{figure}[tbp]
\begin{center}
\scalebox{0.8}{
\includegraphics[width=17.50cm,clip]{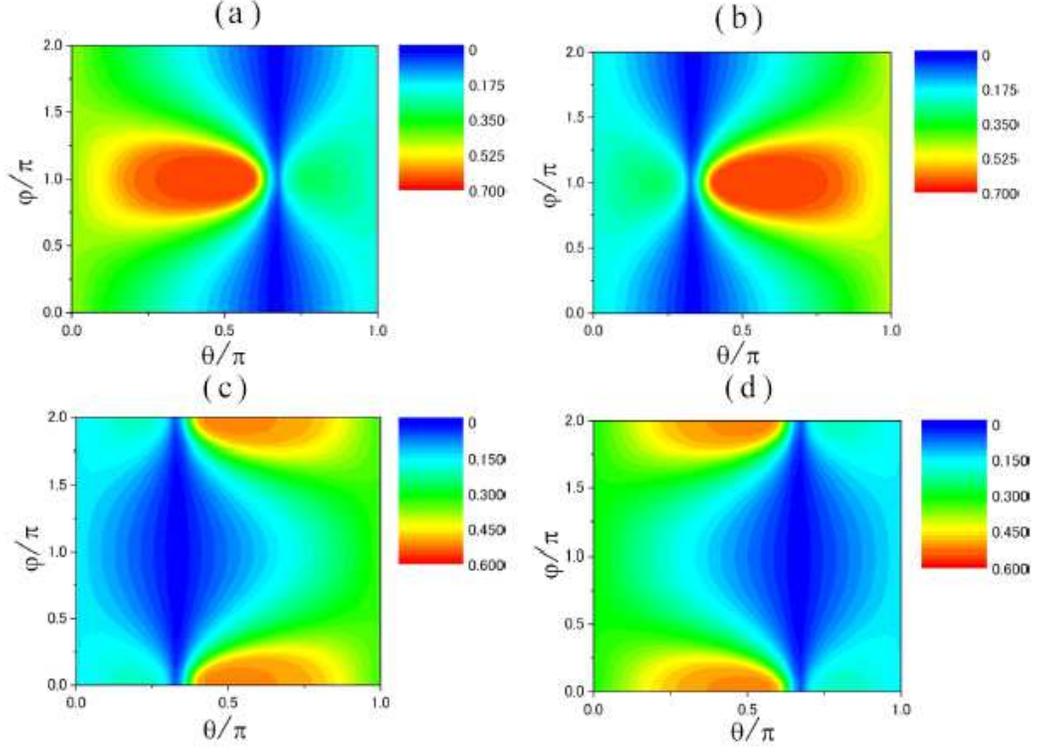}
}
\end{center}
\caption{   Valley resolved conductances (a) $G_{X}$, (b) $G_{X'}$, (c) $G_{Y}$, and (d) $G_{Y'}$ as a function of  the direction of the exchange field $\theta$ and $\varphi$ for $U/E=1$.  We find a valley filtering effect different from that in Fig. 7.}
\label{fig8}
\end{figure}

\begin{figure}[tbp]
\begin{center}
\scalebox{0.8}{
\includegraphics[width=13.50cm,clip]{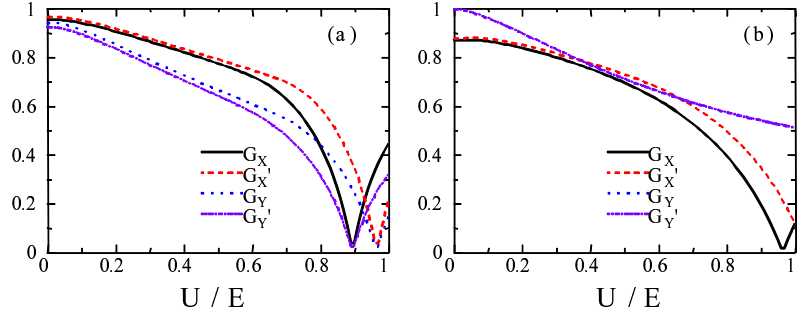}
}
\end{center}
\caption{  Valley resolved conductances $G_{X}$,  $G_{X'}$, $G_{Y}$, and  $G_{Y'}$ as a function of $U/E$. (a) $\theta=\varphi=0$. (b) $\theta=0.5\pi$ and $\varphi=0$. By changing the direction of the magnetization and the potential in the ferromagnetic region, one can control the dominant valley contribution out of four valley degrees of freedom.}
\label{fig9}
\end{figure}

\begin{figure}[tbp]
\begin{center}
\scalebox{0.8}{
\includegraphics[width=13.50cm,clip]{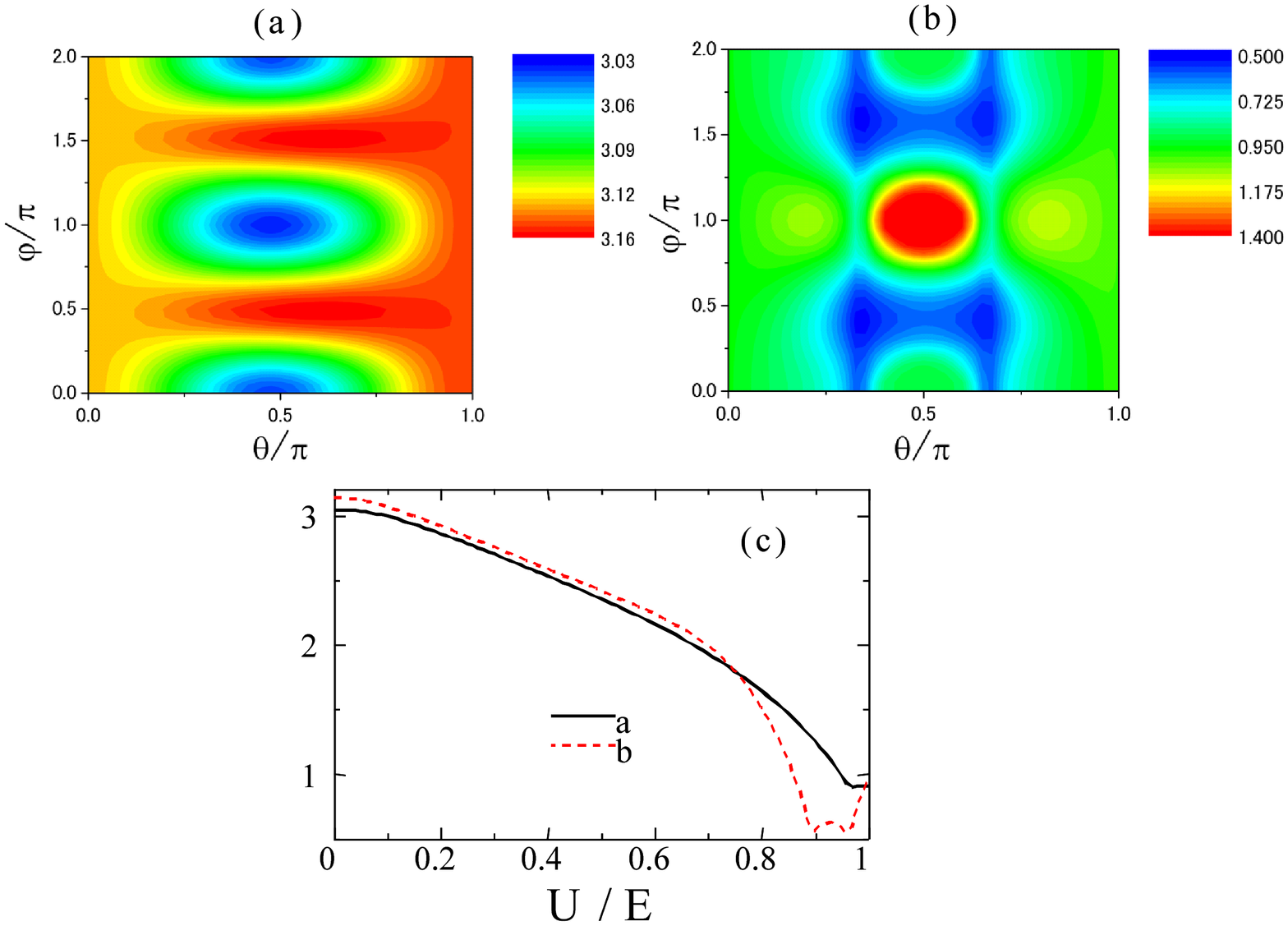}
}
\end{center}
\caption{  Total conductance $G=G_X+G_{X'}+G_{Y}+G_{Y'}$ as functions of  $\theta$ and $\varphi$ for (a) $U=0$ and for (b) $U/E=1$, and as a function of (c) $U$ for  a. $\theta=\varphi=0$, and b. $\theta=0.5\pi$ and $\varphi=0$.}
\label{fig10}
\end{figure}

In the following, we fix $h/E=\lambda _{12} \varepsilon _{12} /E = 0.2,  \lambda _{23} \varepsilon _{23} /E = \lambda _{13} \varepsilon _{13} /E = 0.1,  EL/v_1  = 1$ and $v_2 /v_1  = 0.65$.

Figure \ref{fig7} shows the valley resolved conductances: (a) $G_{X}$, (b) $G_{X'}$, (c) $G_{Y}$, and (d) $G_{Y'}$ as functions of $\theta$ and $\varphi$ for $U=0$. 
As shown in Ref.\cite{Yokoyama4}, the inplane exchange field shifts the Fermi surface in momentum space. The conductance is suppressed due to this shift along $k_y$-direction since $k_y$ is conserved. As for $G_{X}$ and $G_{X'}$, when $h_x$ is positive, the shift of the Fermi surface along the $k_y$-direction is enhanced since $\lambda _{12} \varepsilon _{12} >0$.  Hence, the conductance is strongly suppressed at $\varphi=0$ as seen from Figures \ref{fig7} (a) and (b). On the other hand, when $h_x$ is negative, the shift of the Fermi surface along the $k_y$-direction is canceled. The conductance then becomes large at $\varphi=\pi$. Since the term with $\lambda _{12} \varepsilon _{12}$ in $H_Y$ and $H_{Y'}$ has a sign opposite to that of $h_x$, $G_{Y}$ and $G_{Y'}$ become large at $\varphi=0$ but small at $\varphi=\pi$ as shown in Figures \ref{fig7} (c) and (d).
As $\theta$ deviates from $\pi/2$, the exchange fields points to $z$-direction, and the dependence of the conductances on $\varphi$ becomes weak.
Since $\lambda _{23} \varepsilon _{23} >0$, the mass gap for $H_{X(Y')}$ at $\theta=0$ is larger  than that at $\theta=\pi$. Hence, $G_{X(Y')}$ at $\theta=0$ is smaller than that for $\theta=\pi$. With the same reasoning, we find that $G_{X'(Y)}$ at $\theta=0$ is larger than that for $\theta=\pi$.
It is also seen that in the parameter region where $G_{X}$ and $G_{X'}$ are large, $G_{Y}$ and $G_{Y'}$ can be small and vice versa, indicative of a valley filtering effect. 

In Fig. \ref{fig8}, the valley resolved conductances are plotted as functions of $\theta$ and $\varphi$ for $U/E=1$.
Regarding $G_{X}$ and $G_{X'}$, due to the cancellation of the shift of the Fermi surface, the conductance reaches its maximum at $\varphi=\pi$ as a function of $\varphi$. When $\cos \theta=- \lambda _{23} \varepsilon _{23}/h=-1/2$, namely $\theta=2\pi/3$, $\tilde m$ in $H_X$ becomes zero. Thus, the conductance becomes minimum at this value of $\theta$ as seen in Fig. \ref{fig8} (a). In a similar way, we can understand that $G_{X'}$ and $G_{Y}$ become minimum at $\theta=\pi/3$, while $G_{Y'}$ takes a minimum at $\theta=2\pi/3$. 
In contrast to Fig. \ref{fig7}, in a parameter region with large $G_{X(Y)}$, $G_{X'(Y')}$ can be small and vice versa for $U/E=1$.  
This indicates that the valley filtering effects are controllable by varying the potential in the ferromagnetic region and the direction of the magnetization. 

In Fig. \ref{fig9}, we show valley resolved conductances as a function of $U/E$ for (a) $\theta=\varphi=0$ and (b) $\theta=0.5\pi$ and $\varphi=0$.
In Fig. \ref{fig9} (a), it is found that the relative magnitudes of the conductances depend on $U/E$ which is tunable by gating. 
For $U/E<0.8$, $G_{X}$ and $G_{X'}$ give dominant contributions. Around $U/E=0.9$, $G_{X'}$ and $G_{Y}$ are dominant, while around $U/E=1$, $G_{X}$ shows a dominant contribution.
As shown in Fig. \ref{fig9} (b), $G_{Y}$ and $G_{Y'}$ are dominant contributions around $U/E=1$.
These results indicate that by changing the direction of the magnetization and the potential in the ferromagnetic region, one can control the dominant valley contribution out of four valley degrees of freedom.

We show the total conductance $G$ as functions of $\theta$ and $\varphi$ in Figs. \ref{fig10} (a) and (b), and as a function of $U/E$ in Fig. \ref{fig10} (c). At $U=0$, $G$ takes a maximum around  $\theta=\varphi=0.5\pi$ and a minimum around $\theta=0.5\pi$  and $\varphi=0$ as seen from Fig. \ref{fig10} (a). On the other hand, at $U/E=1$, $G$ takes a maximum for  $\theta=0.5\pi$ and $\varphi=\pi$ as shown in Fig. \ref{fig10} (b). We also have a large magnetoconductance effect compared to the case with $U=0$. Comparing Figs. \ref{fig10} (a) and (b), it is found that by changing $U$, the direction of the magnetization at maximum conductance and that at minimum conductance are exchanged. 
In Fig. \ref{fig10} (c), $G$ is plottd as a function of $U/E$. It is found that the total conductance also depends strongly on the potential in the ferromagnetic region. Therefore, the total conductance is also tunable by electric and magnetic means.

Here, we have considered transport properties on the (001) surface of the topological crystalline insulator. Our formalism is also applicable to the (110) or (111) surfaces of the topological crystalline insulator. 
Recently, angle-resolved photoemission spectroscopy on the (111) surface of the topological crystalline insulator has been reported. Dirac cones at the $\bar{\Gamma}$ and $\bar{M}$ points have been observed.\cite{Tanaka3,Polley} It has been also revealed that the energy location of the Dirac point and the Dirac velocity are different at the $\bar{\Gamma}$ and $\bar{M}$ points.\cite{Tanaka3}
These characteristics can be taken into account in our formalism by changing parameters $v_1$, $v_2$ and $U$ at each valley.

\section{Conclusions}

In summary,
we have investigated spin and valley transports in junctions composed of silicene and topological crystalline insulators. 
We have considered normal/magnetic/normal Dirac metal junctions where a gate electrode is attached to the magnetic region. 
In normal/antiferromagnetic/normal silicene junction, it is shown that the current through this junction is valley and spin polarized due to the coupling between valley and spin degrees of freedom, and the valley and spin polarizations can be tuned by local application of a gate voltage. In particular, we have found a fully valley and spin polarized current by applying the electric field. 
In normal/ferromagnetic/normal topological crystalline insulator junction with a strain induced in the ferromagnetic segment, we have investigated valley resolved conductances and clarified how the valley polarization stemming from the strain and exchange field appears in this junction. It is found that changing the direction of the magnetization and the potential in the ferromagnetic region, one can control the dominant valley contribution out of four valley degrees of freedom.
We have also reviewed spin transport in  normal/ferromagnetic/normal graphene junctions, and spin and valley transports in normal/ferromagnetic/normal silicene junctions.

The role of magnetism is different in graphene, silicene and topological crystalline insulator junctions.
In graphene junctions, the ferromagnetism induces different chemical potential shifts for up and down spin states, which leads to the shift of the oscillation of the conductances. As a result, a finite spin current appears.
In silicene junctions, the (anti)ferromagnetism opens different spin dependent band gaps at $K$ and $K'$ points. This results in spin and valley polarized transports in these junctions. 
In topological crystalline insulator junctions, the ferromagnetism also induces  valley dependent band gaps and inplane ``vector potentials" in combination with strain effects. These properties lead to valley dependent transports.

Note added: Recently, we learned of a related work on the ferromagnetic silicene junctions.\cite{Soodchomshom}

The author thanks S. Murakami, Y. Okada, M. Ezawa, and X. Hu for helpful comments and discussions.
This work was supported by Grant-in-Aid for Young Scientists (B) (No. 23740236) and the ``Topological Quantum Phenomena" (No. 25103709) Grant-in Aid for Scientific Research on Innovative Areas from the Ministry of Education, Culture, Sports, Science and Technology (MEXT) of Japan.

\end{document}